\documentclass[aps,prstab,twocolumn,superscriptaddress,showpacs,showkeys]{revtex4-1}

\usepackage[pdftex]{graphicx}
\usepackage[upright]{fourier} % for non-slanted greek letters
\usepackage{amsmath,color,float,hyperref,scrextend}

\begin{document}

%\preprint{}

\title{Design and Vertical Tests of SPS-series Double-Quarter Wave (DQW) Cavity Prototypes for the HL-LHC Crab Cavity System}

% Please use the appropriate macro for each type of information

%\thanks{}


\author{S. Verd\'u-Andr\'es}
\email[Corresponding author: ]{sverdu@bnl.gov}
\affiliation{Brookhaven National Laboratory BNL, Upton, NY 11973, United States of America}

\author{K. Artoos}
\affiliation{European Organization for Nuclear Research CERN, 1217 Meyrin, Switzerland}

\author{S. Belomestnykh}
\affiliation{Brookhaven National Laboratory BNL, Upton, NY 11973, United States of America}
\affiliation{Fermi National Laboratory, Batavia, IL 60510, United States of America}
\affiliation{Stony Brook University, Stony Brook, NY 11790, United States of America}

\author{I. Ben-Zvi}
\affiliation{Brookhaven National Laboratory BNL, Upton, NY 11973, United States of America}
\affiliation{Stony Brook University, Stony Brook, NY 11790, United States of America}

\author{C. Boulware}\affiliation{Niowave Inc., East Lansing, MI 48906, United States of America}

\author{G. Burt}
\affiliation{Lancaster University, Bailrigg, Lancaster LA1 4YW, United Kingdom}
\affiliation{Cockcroft Institute, Daresbury, Warrington WA4 4AD, United Kingdom} 

\author{R. Calaga}
\affiliation{European Organization for Nuclear Research CERN, 1217 Meyrin, Switzerland}

\author{O. Capatina}
\affiliation{European Organization for Nuclear Research CERN, 1217 Meyrin, Switzerland}

\author{F. Carra}
\affiliation{European Organization for Nuclear Research CERN, 1217 Meyrin, Switzerland}

\author{A. Castilla}
\affiliation{European Organization for Nuclear Research CERN, 1217 Meyrin, Switzerland}

\author{W. Clemens}\affiliation{Jefferson Lab, Newport News, VA 23606, United States of America}

\author{T. Grimm}\affiliation{Niowave Inc., East Lansing, MI 48906, United States of America}

\author{N. Kuder}
\affiliation{European Organization for Nuclear Research CERN, 1217 Meyrin, Switzerland}

\author{R. Leuxe}
\affiliation{European Organization for Nuclear Research CERN, 1217 Meyrin, Switzerland}

\author{Z. Li}
\affiliation{SLAC National Accelerator Laboratory, Palo Alto, CA 94025, United States of America}

\author{E. A. McEwen}\affiliation{Jefferson Lab, Newport News, VA 23606, United States of America}

\author{H. Park}\affiliation{Jefferson Lab, Newport News, VA 23606, United States of America}

\author{T. Powers}\affiliation{Jefferson Lab, Newport News, VA 23606, United States of America}

\author{A. Ratti}\affiliation{SLAC National Accelerator Laboratory, Palo Alto, CA 94025, United States of America}\affiliation{Lawrence Berkeley National Laboratory, Berkeley, CA 94720, United States of America}

\author{N. Shipman}
\affiliation{Lancaster University, Bailrigg, Lancaster LA1 4YW, United Kingdom}
\affiliation{Cockcroft Institute, Daresbury, Warrington WA4 4AD, United Kingdom}

\author{J. Skaritka}
\affiliation{Brookhaven National Laboratory BNL, Upton, NY 11973, United States of America}

\author{Q. Wu}
\affiliation{Brookhaven National Laboratory BNL, Upton, NY 11973, United States of America}

\author{B. P. Xiao}
\affiliation{Brookhaven National Laboratory BNL, Upton, NY 11973, United States of America}

\author{J. Yancey}\affiliation{Niowave Inc., East Lansing, MI 48906, United States of America}

\author{C. Zanoni}
\email[Now at: European Southern Observatory (ESO), Munich, Germany.]{}
\affiliation{European Organization for Nuclear Research CERN, 1217 Meyrin, Switzerland}

\date{\today}

\begin{abstract}
Crab crossing is essential for high-luminosity colliders. The High Luminosity Large Hadron Collider (HL-LHC) will equip one of its Interaction Points (IP1) with Double-Quarter Wave (DQW) crab cavities. A DQW cavity is a new generation of deflecting RF cavities that stands out for its compactness and broad frequency separation between fundamental and first high-order modes. The deflecting kick is provided by its fundamental mode. Each HL-LHC DQW cavity shall provide a nominal deflecting voltage of 3.4~MV, although up to 5.0~MV may be required.  A Proof-of-Principle (PoP) DQW cavity was limited by quench at 4.6~MV. This paper describes a new, highly optimized cavity, designated DQW SPS-series, which satisfies dimensional, cryogenic, manufacturing and impedance requirements for beam tests at SPS and operation in LHC. Two prototypes of this DQW SPS-series were fabricated by US industry and cold tested after following conventional SRF surface treatment. Both units outperformed the PoP cavity, reaching a deflecting voltage of 5.3--5.9~MV. This voltage -- the highest reached by a DQW cavity -- is well beyond the nominal voltage of 3.4~MV and may even operate at the ultimate voltage of 5.0~MV with sufficient margin. This paper covers fabrication,  surface preparation and cryogenic RF test results and implications.   
\end{abstract}


\maketitle


\section{Introduction}
Crab crossing is an essential mechanism for high-luminosity colliders. The High Luminosity Large Hadron Collider (HL-LHC) will implement crab crossing at the Interaction Point (IP) of ATLAS (IP1) and CMS (IP5)~\cite{bib:Calaga15}.  The crabbing system of HL-LHC will follow the local scheme~\cite{bib:Calaga15}. The IP1 will be equipped with a set of Double-Quarter Wave (DQW) cavities~\cite{bib:DQW} while IP5 will have a set of RF Dipole (RFD) cavities~\cite{bib:RFD}. The four-rod crab cavity~\cite{bib:Hall17} was also considered for the HL-LHC crabbing system and then downselected in favor of the DQW and RFD cavities. Both DQW and RFD will be Superconducting Radio-Frequency (SRF) bulk niobium cavities and will operate at 400~MHz in Continuous Wave (CW) mode ~\cite{bib:TDR}. Crab crossing also appears in the baseline design of electron-ion colliders eRHIC and JLeIC. Both will require crab crossing to reach the necessary luminosity levels for relevant nuclear physics studies~\cite{bib:eRHICpCDR, bib:JLeIC}. The eRHIC crab cavities are based on the DQW design developed for HL-LHC~\cite{bib:VerduAndres17}. 

The operational experience of crab cavities with beam is currently limited to the electorn-positron collider KEK-B~\cite{bib:Funakoshi14}. Electron-positron beams are highly damped by synchrotron radiation, thus more robust against crabbing errors than proton machines. The KEK-B crab cavities were elliptical. With a higher-order crabbing mode, their size was too large to fit between the closely spaced beam pipes of the LHC. This led to the development of the compact DQW and RFD cavities. Before crabbing the proton bunches of LHC, a cryomodule with two fully-dressed DQW cavities will be tested with beam in SPS in 2018 to address crab cavity operational issues in a proton machine. Two DQW SPS-series cavity prototypes were fabricated in US industry following conventional surface treatments for SRF cavities. The two DQW cavities and equipment used for the SPS tests were fabricated in-house by CERN~\cite{bib:Zanoni17}. All four cavities share the same RF design. 

The DQW cavity belongs to a new generation of deflecting RF cavities.  The fundamental mode of a DQW cavity provides the necessary deflecting kick for bunch crabbing. The DQW cavity stands out for its compactness and broad frequency separation between fundamental and first high-order modes. The DQW cavity can be seen as two coaxial Quarter Wave (QW) resonators mirror symmetric with respect to their open-end plane. The opposing inner conductor poles behave as capacitor plates. The highest magnetic field region of the fundamental mode is found in the cavity dome (the shorted ends of the QW resonator), whereas the highest electric field region is in the capacitive plates. The field distribution in the rest of the cavity body corresponds to a TEM-like mode. The voltage sustained between the plates provides a deflecting kick to the bunch. The accelerating voltage seen by a bunch traveling on axis through a symmetric DQW is essentially zero. 

\begin{figure}[!htb]
   \centering
  \includegraphics*[width=240pt]{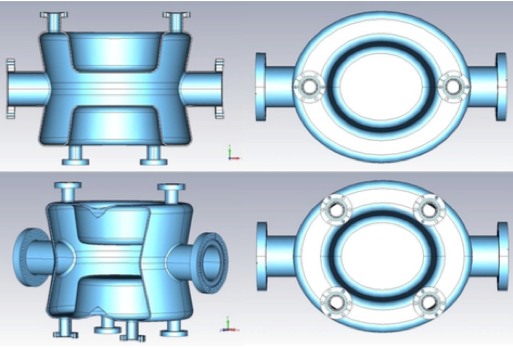}
   \caption{\label{fig:PoP-scheme}Views of PoP DQW cavity~\cite{bib:Xiao13}.}
\end{figure}

The current HL-LHC baseline accounts for a total of 16 crab cavities: 2 cavities per IP per side per beam~\cite{bib:Rossi16}. The crab cavities of HL-LHC will be located between dipole D2 and quadrupole Q4 in the LHC Interaction Region (IR). In this location, the two beams are well separated into their corresponding individual pipes and the betatron function is large enough to minimize the required crabbing voltage~\cite{bib:Calaga15}. Each cavity will provide a nominal deflecting voltage of 3.4~MV, but up to 5.0~MV may be required for full geometric overlap of the colliding bunches. The delivery of such high deflecting voltage requires operation at significantly high peak surface fields.  A Proof-of-Principle (PoP) DQW cavity was fabricated by Niowave Inc. in 2013 for RF performance validation. The PoP cavity reached a peak surface magnetic field of 116 mT before quenching at the maximum deflecting voltage of 4.6 MV~\cite{bib:Xiao15}. Views of the PoP cavity are shown in Fig.~\ref{fig:PoP-scheme}. A new, highly optimized version -- designated the DQW SPS-series --  satisfies dimensional, cryogenic, manufacturing and impedance requirements for operation in the LHC. This new design also presents lower peak fields with the aim at reaching higher deflecting voltages than the PoP cavity. 

The present paper is organized as follows. The first part of the paper discusses the main design features of the SPS-series cavities, and provides comparisons with the PoP cavity. The second part describes fabrication and surface treatment of two US DQW SPS-series cavity prototypes built to test the cavity concept and steer the development of the SPS cavities at CERN. The third part presents bare cavity cold test results and discusses cryogenic RF performances of these prototypes. Fabrication and test results of the CERN cavities are out of the scope of this paper.  

\section{SPS-series DQW Cavity Design}
\subsection{RF design}
\begin{figure}[!htb]
   \centering
  \includegraphics*[width=240pt]{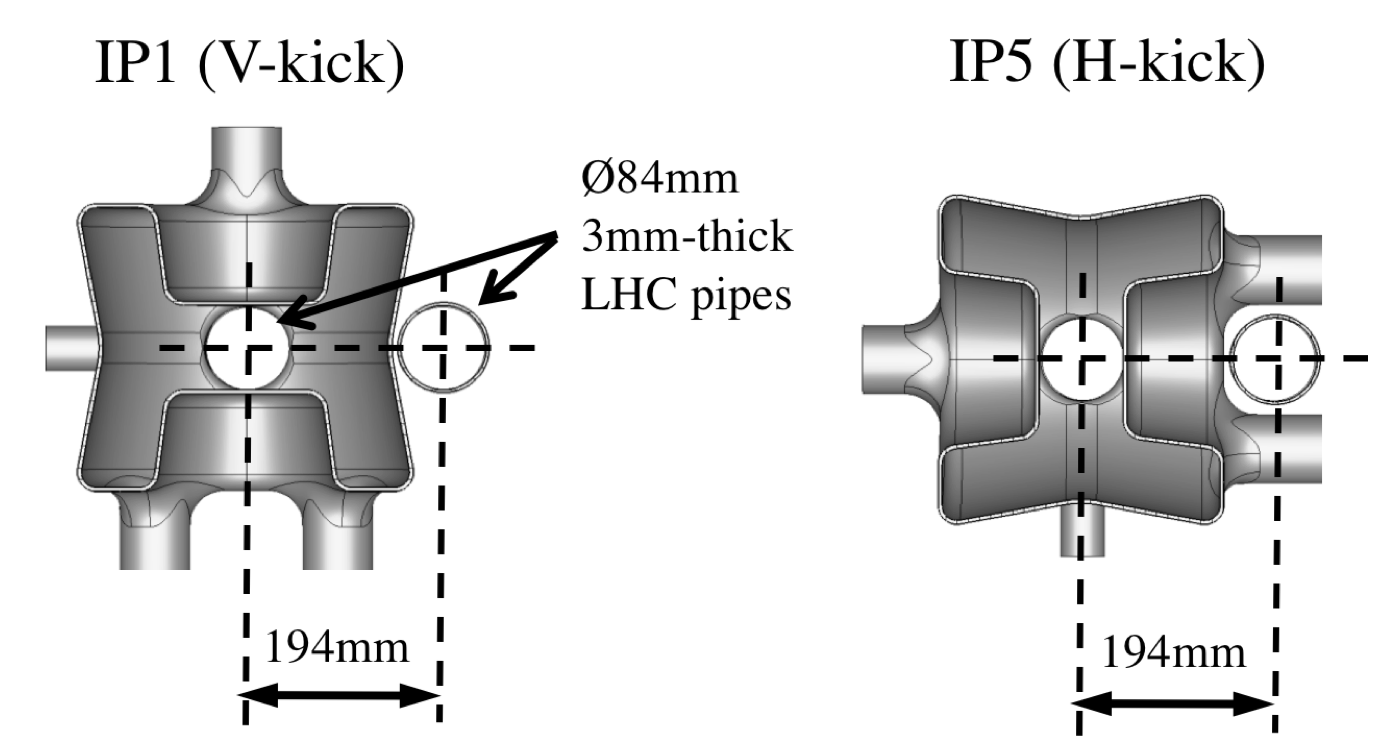}
   \caption{\label{fig:VH} The SPS-series DQW cavity satisfies the LHC geometric restrictions to provide a deflecting kick in both vertical and horizontal kick configurations.}
\end{figure}

The SPS-series DQW cavity presents three main important new features. First of all, the design is compatible with both vertical and horizontal kick configurations. Crab crossing will be implemented for two LHC IPs: IP1 (ATLAS) and IP5 (CMS). The beams cross in the vertical plane at IP1 and in the horizontal plane at IP5. The second beam pipe of the LHC limits the cavity width (when used in vertical kick configuration) and height (when used in horizontal kick configuration)~\cite{bib:Baudrenghien13}. Similarly to the PoP cavity design, the cavity adopts a pronounced hourglass shape to accommodate the second beam pipe in IP1.  In addition, the cavity takes an elliptical racetrack profile for frequency tuning as the cavity height is fixed by the second beam pipe in IP5. Fig.~\ref{fig:VH} shows the DQW crab cavity for both vertical and horizontal kick configurations. The slimmer section is usually referred to as cavity waist. 

\begin{figure}[h]
   \centering
  \includegraphics*[width=192pt]{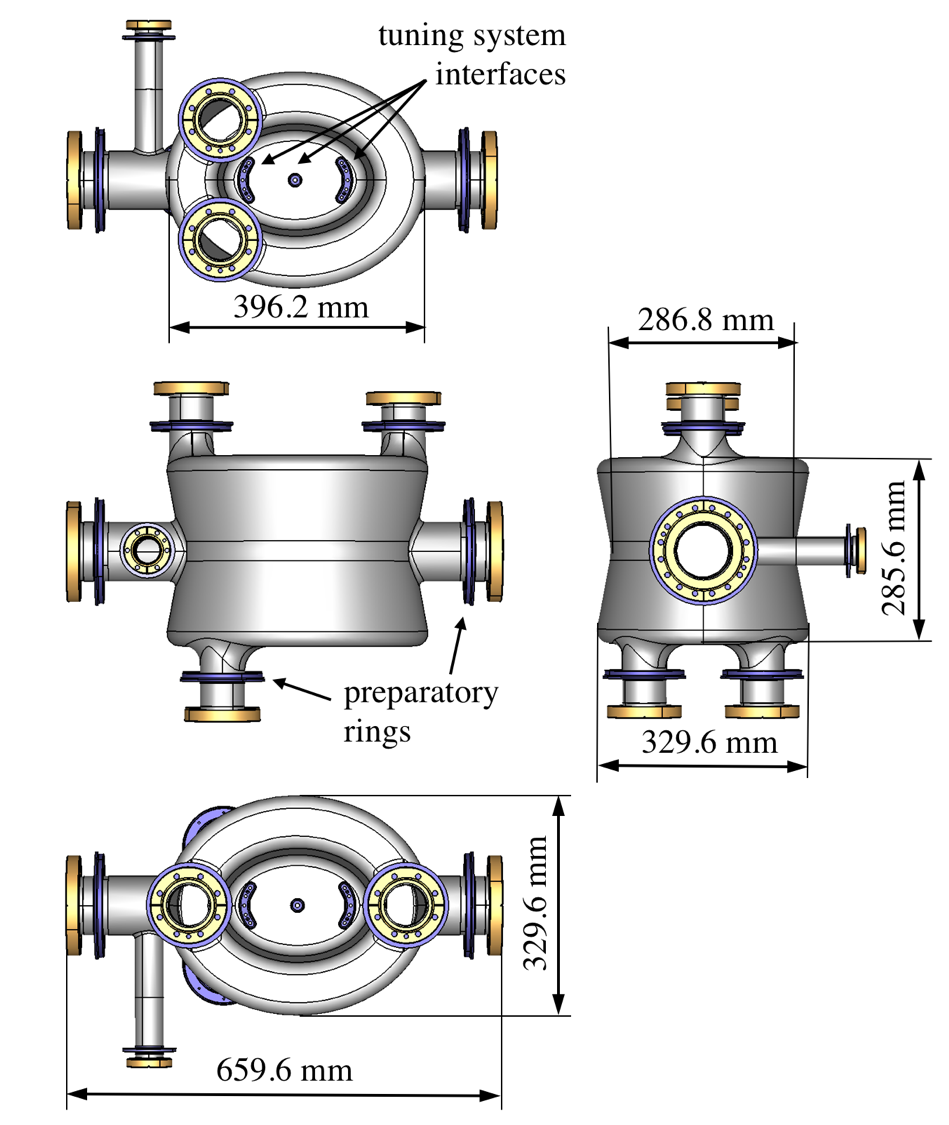}
   \caption{\label{fig:PortConf}Main dimensions at warm of SPS-series DQW cavity mechanical design.}
\end{figure}

Second, the RF coupling ports are large enough to incorporate a 40~kW Fundamental Power Coupler (FPC) and ensure the extraction of 1~kW Higher Order Mode (HOM) power with a minimum number of ports~\cite{bib:Calaga15, bib:XiaoSRF13}. In total, the cavity has four 62-mm-diameter ports. One port serves the FPC; the other three host the HOM filters. To ease cleaning of the cavity's interior, these four ports are located in the cavity's dome -- the inductive region of the cavity. For efficient coupling to the magnetic field, hook-type coupling elements are used for both the FPC and HOM filters. The hook also provides good coupling to the two lowest HOMs, found at 567~MHz and 588~MHz, both with high magnetic field concentrated in the cavity dome.  All couplers are detachable from the cavity to facilitate cleaning and installation. The elliptical racetrack has a constant width to accommodate the port openings. The selected port configuration, shown in Fig.~\ref{fig:PortConf}, provides the lowest external Q of HOMs for frequencies of up to 2~GHz~\cite{bib:Wu13}. The modes with frequencies above 2~GHz are expected to be Landau damped~\cite{bib:Baudrenghien13}. Two ports are in line with the beam axis; the other two are at an angle of 45~degrees with respect to the beam axis to allow the passage of the second beam pipe of LHC when the cavity is used to provide a horizontal kick. The port distribution asymmetry introduces a residual accelerating voltage of 15~kV (for a nominal deflecting voltage of 3.4~MV) and an electric field center offset of 0.23~mm (displaced towards the cavity bottom). The pickup port, opened on one of the beam pipes to preserve the cavity's symmetry, is perpendicular to the deflecting kick direction to ease cavity manufacturing. A DQW SPS-series cavity with three HOM filters and the pickup meets the impedance budget imposed for operation in the LHC~\cite{bib:Biancacci16}. Detailed discussion of the design, fabrication and performance of the HOM filters and pickup will be addressed in a separate communication. 

Finally, the port-cavity interface is optimized to reduce the magnetic peak field~\cite{bib:VerduAndres13}. The magnetic field finds its maximum close to the port apertures in the cavity body. For a bare cavity with no port openings, a maximum peak magnetic field of 66~mT (for a nominal deflecting voltage of 3.4~MV) is found along the inner radius of the cavity dome. Design of the port-cavity interface aimed to achieve a geometry with reduced field enhancement that was also easy to fabricate. The selected port-cavity interface leads to a maximum peak surface magnetic field of 73~mT for a nominal deflecting voltage of 3.4~MV.  The highest field is located between the 45-degree ports at the inner radius of the cavity dome. 

The main electromagnetic properties of the SPS-series DQW cavity are given in Table~\ref{tab:EM}. The shunt impedance of the deflecting mode $R_{\textrm{t}}/Q$ is defined following the accelerator convention as: 

\begin{equation}
\frac{R_{\textrm{t}}}{Q} = \frac{V_{\textrm{t}}^{2}}{\omega U}
\label{eq:RtQ}
\end{equation}

where $U$ is the stored energy. Note that $R_{\textrm{t}}/Q$ is given in units of impedance, instead of impedance per unit length, as is typically the convention for transverse shunt impedance. The SPS-series DQW cavity has an $R_{\textrm{t}}/Q$ of 429.3~$\Omega$ for the fundamental (crabbing) mode. The geometry factor $G$, defined as the product of quality factor $Q_{0}$ and surface resistance $R_{\textrm{s}}$, is 87 $\Omega$. This is a low geometry factor compared to the typical TESLA-type elliptical cavity ($G \sim 270~\Omega$)~\cite{bib:TESLA}. Since the BCS surface resistance for niobium at 400~MHz and 2~K is only 1~n$\Omega$, the residual surface resistance has a significant impact on the final $Q_{0}$ that the cavity can achieve. For a surface resistance of 10~n$\Omega$, the maximum quality factor of the cavity at low fields will be $8.7\times 10^{9}$.   The dynamic power dissipated per cavity specified for tests in the SPS and operation in the LHC is about 5~W at the operating voltage of 3.4 MV~\cite{bib:TDR}. 

Table~\ref{tab:Geo} summarizes the main geometry properties of the SPS-series DQW crab cavity. General cavity dimensions are displayed in Fig.~\ref{fig:PortConf}. 

\begin{figure}[tb]
   \centering
  \includegraphics*[width=250pt]{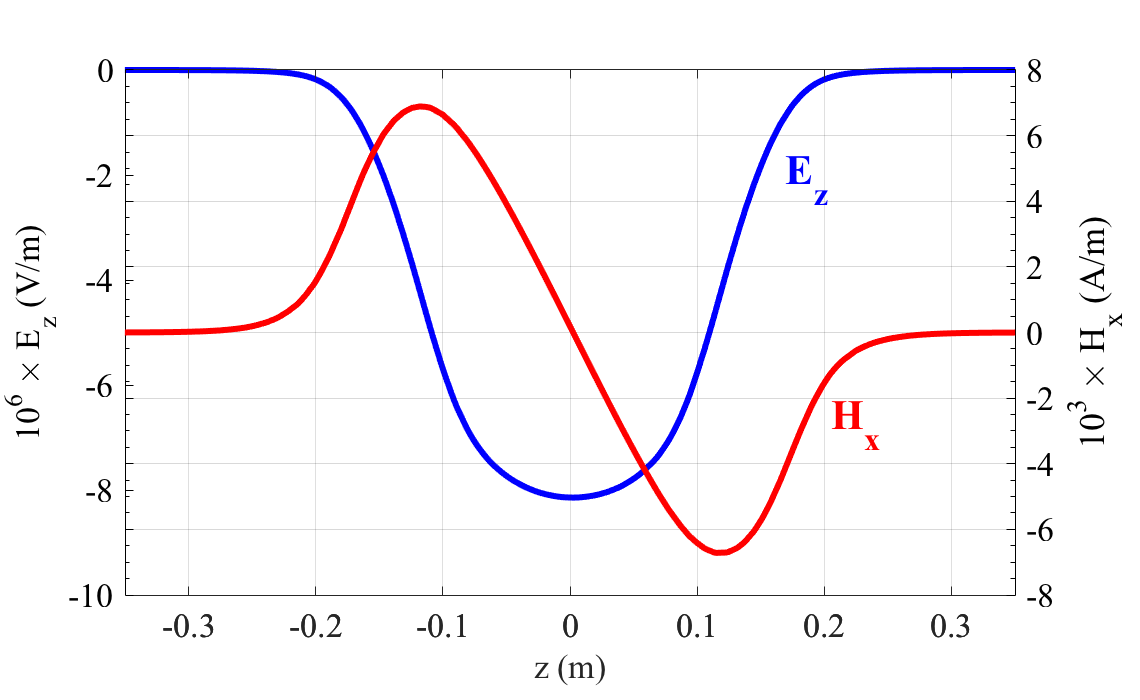} 
   \caption{ \label{fig:EandH} Electric and magnetic field along the geometric center of a DQW cavity.}
\end{figure} 

\begin{table}[h]
   \centering
   \caption{\label{tab:Geo}General geometry dimensions of the SPS-series DQW cavity at ambient temperature (includes wall thickness).}
\begin{ruledtabular}
\begin{tabular}{lcc}
\ Cavity length (flange to flange) &  659.60 & mm\\
\ Cavity height (port to port) &  495.71& mm\\
\ Cavity width (no pickup) & 329.60& mm \\
\ Beam pipe inner diameter  & 84 & mm\\
\ FPC and HOM port inner diameter & 62& mm\\
\end{tabular}
\end{ruledtabular}
\end{table}

\begin{table}[h]
   \centering
   \caption{\label{tab:EM}Electromagnetic properties of the SPS-series DQW cavity.}
\begin{ruledtabular}
 \begin{tabular}{lccl}
\ Fundamental frequency
& f$^{(0)}$ & 400.79 & MHz \\
\ First HOM frequency (longitudinal mode)& 
f$^{(1)}$ & 567 & MHz \\
\ Transverse R/Q & 
R$_{\textrm{t}}$/Q & 429.3 & Ohm \\
\ Geometry factor &
G & 87 & Ohm \\
\ Peak surface magnetic field\footnote[1]{\label{lab:first}For nominal deflecting voltage of 3.4 MV.}& $B_{\textrm{p}}$ & 72.8 & mT \\
\ Peak surface electric field\footref{lab:first} & $E_{\textrm{p}}$ & 37.6 & MV/m \\
\ Residual accelerating voltage\footref{lab:first} & $V_{\textrm{acc}}$ & 15.23 & kV \\
\ Electric field center offset & $O_{E}$
& -0.23 & mm \\
\end{tabular}
\end{ruledtabular}
\end{table}

\subsection{Multipacting}
The phenomenon of multipacting can be a serious obstacle for normal operation of RF cavities. Multipacting currents absorb RF power, degrade the coupling between power source and RF cavity, heat the impacted surfaces to the extent of causing thermal breakdown in SRF cavities and even break ceramic windows.  

\begin{figure}[!htb]
   \centering
  \includegraphics*[width=240pt]{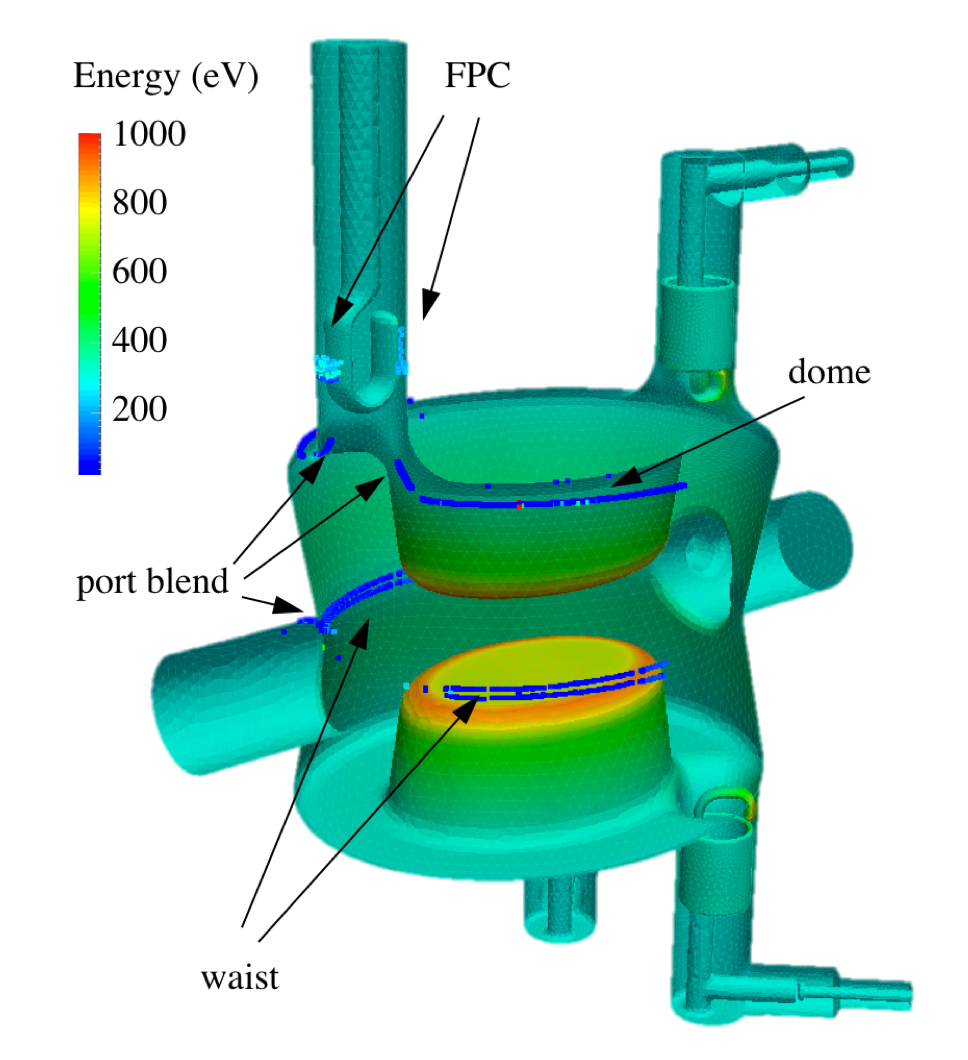}
   \caption{\label{fig:MP}Multipacting sites found in a DQW cavity by ACE3P/Track3P. Color scale represents the impact energy in eV of multipacting electrons.}
\end{figure}

Simulations were conducted with the particle tracking codes Track3P of ACE3P~\cite{bib:ACE3P} and Particle Studio of CST~\cite{bib:CST} to identify the potential multipacting sites and find at which voltage levels multipacting may occur. The Secondary Emission Yield (SEY) curve for baked niobium was used in our studies~\cite{bib:Calder86}. Fig.~\ref{fig:MP} displays the predicted multipacting sites by ACE3P in a SPS DQW cavity. Multipacting bands are found in: 1) the cavity waist for deflecting voltage levels below 0.5~MV; 2) in the blending of the cavity dome, between 2 and 3~MV and between 4.0 and 4.5~MV; 3) in the blending of the beam ports, between 1.6 and 4.0~MV; 4) the blending of the small cavity ports, between 1.7 and 2.7~MV and; 5) in the FPC port, below 0.5~MV. CST predicted similar multipacting bands and also identified a multipacting band between 1 and 2.5 MV in the cavity waist.  

\begin{figure}[!htb]
   \centering
 	\includegraphics*[width=210pt]{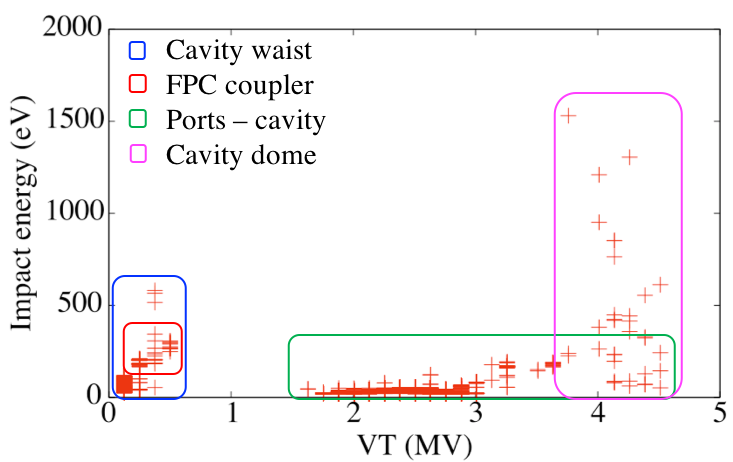}
   \caption{\label{fig:MP2} Impact energy of multipacting electrons found by ACE3P/Track3P simulations at different deflecting voltage levels.}
\end{figure}

Multipacting in the FPC port may require dedicated high-power conditioning. The maximum SEY of baked niobium is above 1 for  for electrons with impact energy between 100 and 1500 eV, with the maximum SEY found at around 300 eV.  Fig.~\ref{fig:MP2} shows the impact energy of multipacting electrons at different deflecting voltage levels. Multipacting in the HOM filters will be discussed in a future publication. 

As a reference, we may consider the tests of the PoP DQW cavity, with a structure similar to the SPS DQW cavity of this report. The PoP DQW cavity exhibited multipacting below 0.5~MV and between 2 and 3~MV. Low field multipacting could be quickly processed through during the first test of the PoP cavity and was not seen again in later cold tests at BNL~\cite{bib:Xiao15}.

\subsection{Mechanical design}
 
The position of the FPC port flange was chosen as a compromise between the losses in the RF-seal copper gasket and the conductive losses in the FPC tube. The length of the HOM ports is determined by the location of the capacitive cylinder in the HOM filter, which rejects most of the fundamental mode power back. The detailed HOM filter design is shown in Ref.~\cite{bib:XiaoIPAC15}. The port is cut beyond the position of this capacitive cylinder. The largest power loss, 0.1~W, is found for the RF-seal gasket in the shortest beam pipe. The loss in the RF-seal gaskets of the four 62 mm-diameter ports only amounts to 23~mW. The loss in the FPC tube is below 1~mW. All these values are computed for a nominal deflecting voltage of 3.4~MV.  

The cavity prototypes include the interfaces to helium vessel and tuning system. Fig.~\ref{fig:PortConf} shows a bare DQW crab cavity with flanges, preparatory rings and interfaces to the tuning system. 

\section{Cavity fabrication}
\begin{figure}[!htb]
   \centering
  \includegraphics*[width=163pt]{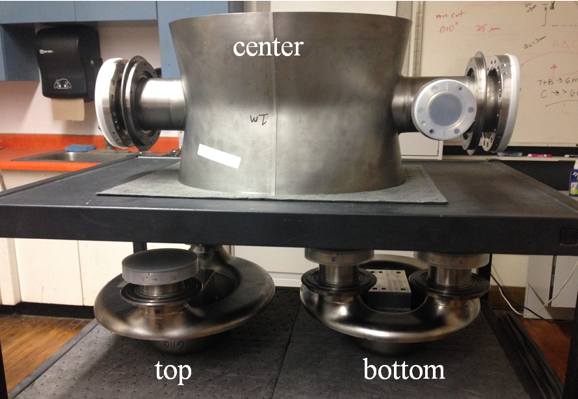}
  \includegraphics*[width=80pt]{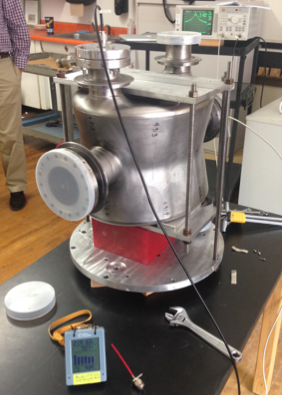}
   \caption{\label{fig:Sub}[Left] Subassemblies for a DQW cavity. [Right] Clamped assembly of a DQW cavity.}
\end{figure}

The fabrication model was prepared considering the volume changes due to BCP,  cool-down and weld-shrinkage. The parts were deep drawn from fine grain RRR $\geq$ 300 niobium sheets and then electron beam welded to form the three main subassemblies shown in Fig.~\ref{fig:Sub}. The main cavity body parts were fabricated out of 4~mm-thick sheets to withstand a pressure difference of 1.8~bar on the cavity walls (2.7~bar after application of the safety coefficient) before yielding~\cite{bib:Kuder}. The extrusion of port nipples inevitably resulted in a local reduction of the material thickness. The cavity extremities were manufactured from 3 mm-thick sheets to connect with the extruded port nipples. The cavity ports were equipped with 90-degree knife-edge CF flanges. All flanges were made from multidirectional (3D) forged, austenitic stainless steel grade 316~LN. The stainless-steel flanges were vacuum brazed to the niobium tubes. The ports also incorporated NbTi adaptors to interface the niobium cavity with the titanium helium vessel~\cite{bib:Zanoni15}. 

The installed cavity, operating at 2~K and delivering 3.4~MV to the 450~GeV proton beam of SPS, will operate at (400.79 $\pm$ 0.06)~MHz. The target frequency for the manufactured cavity was 400.29~MHz at ambient conditions, after considering the frequency changes due to BCP, coupler insertion, evacuation and cool-down. 

\begin{figure}[!htb]
   \centering
 \includegraphics*[width=250pt]{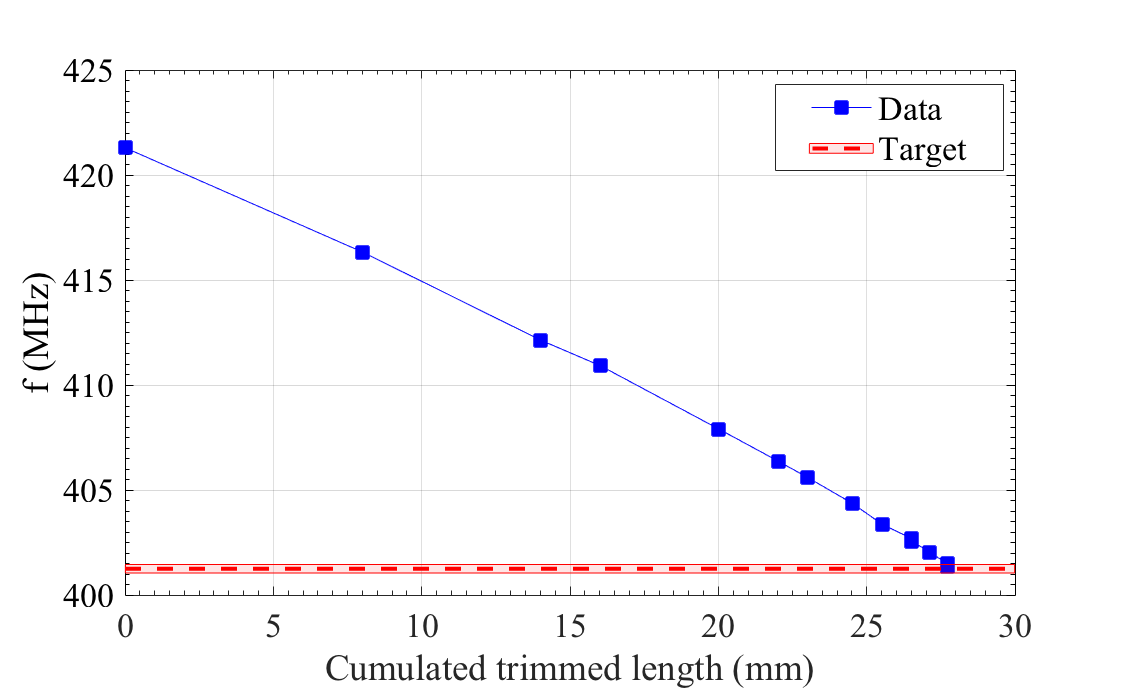}
   \caption{ \label{fig:TrimTuning}Frequency evolution during trim tuning of cavity \#1. Each data point corresponds to the frequency measured at a certain trimming stage. The dashed red line represents the target frequency for the clamped assembly, 401.25~MHz, while the red shadowed area represents the frequency tolerance, $\pm$0.20~MHz.} % ({\color{blue}{$\blacksquare$}})
\end{figure}

The first tuning of the cavities took place during cavity fabrication, when the cavity was still divided into three main subassemblies. The subassemblies were clamped together and the frequency of the assembly was measured. Then, some material was trimmed out of the subassembly edges to reduce the cavity height and thus tune the cavity frequency. The material removal was performed in several iterations to carefully approach the target frequency. Equal amount of material was trimmed at one side and the other of the beam axis to preserve the alignment of the electric field center. Fig.~\ref{fig:TrimTuning} shows the frequency evolution for the clamped assembly of cavity \#1 as material is removed from the subassembly edges. 

Before the last trimming step, the wall was thinned down to 3~mm along the seams of the two last welds. The thinning intended to improve the matching of subassembly edges, ease weld execution and reduce the weld penetration depth. This operation had already been implemented in the past for other cavities~\cite{bib:Sears05}. ANSYS simulations~\cite{bib:ANSYS} found that this local thickness reduction should not compromise the structural integrity of the DQW cavity during cooldown~\cite{bib:Kuder}. A small amount of material was removed from the internal surface (or RF surface) while most of the material was removed from the outer surface. The interior surface was machined with a ball mill to ensure a smooth transition. 

Upon reception at Jefferson Lab, the frequency of the clamped assembly was 401.35~MHz for cavity \#1, within the accepted range of (401.25 $\pm$0.20)~MHz. The trim tuning process demonstrated the required control for cavity production. The cavity subassemblies were then joined using electron beam welding at Jefferson Lab. The welding was expected to reduce the assembly frequency by almost 1 MHz. However, the welding process shifted the frequency of cavity \#1 in the opposite direction, to 401.60~MHz. The frequency of cavity \#2 also did not behave as expected upon welding. Later work on the CERN cavities found that the cavities were deformed by the two last welds. Such deformation had not been considered in the frequency shift estimation. A tuning method was implemented at CERN to revert the deformation and tune the cavities to the target frequency~\cite{bib:Verdu17}.

\section{Surface preparation and cavity assembly} 
Surface preparation and cold tests of the two cavities were conducted in the SRF facility of Jefferson Lab. The cavity surface was prepared following the Jefferson Lab standard treatment procedures: ultrasound bath degreasing, bulk Buffered Chemical Polishing (BCP), 10 hour hydrogen degassing in UHV furnace at 600$^{\circ}$C, another ultrasound bath degreasing, light BCP, manual rinsing of every port followed by High-Pressure Rinsing (HPR) in dedicated HPR cabinet with ultra-pure water (cavity in vertical orientation, sprinkler moving up and down along beam pipe axis), assembly in class 10 (ISO 4) clean-room, slow bleed pumping, leak check and 120$^{\circ}$C baking of the evacuated cavity in furnace dedicated to SRF applications.  Dimensional and frequency controls were done at different stages of the surface treatment. 

\begin{figure}[!htb]
   \centering
  \includegraphics*[width=240pt]{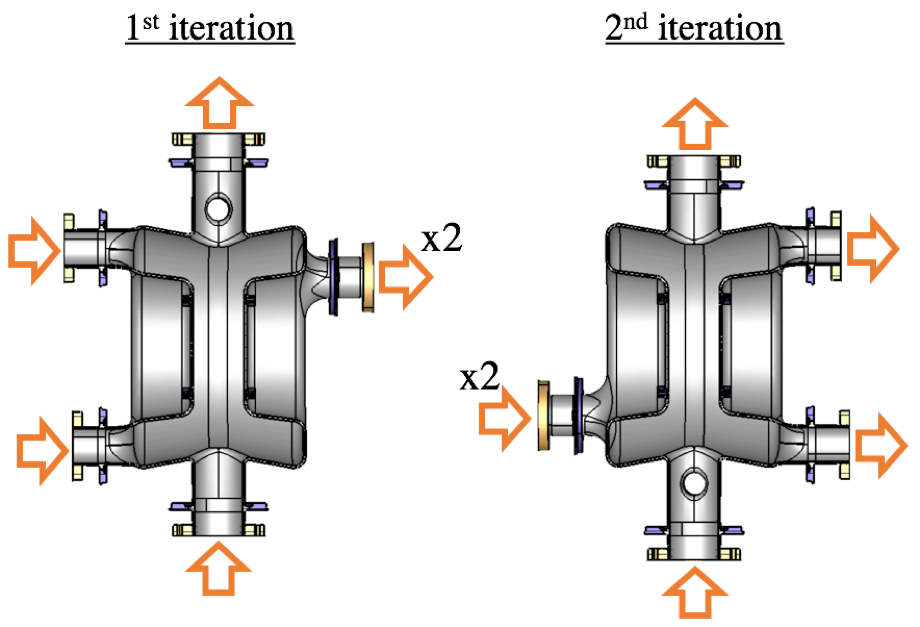}
   \caption{ \label{fig:BCP}Bulk BCP of cavity in 2 iterations.}
\end{figure}

The BCP treatment was conducted in a fixed bench using an acid mixture of HF (48\% conc.), HNO$_{3}$ (70\% conc.) and H$_{3}$PO$_{4}$ (85\% conc.) in a 1:1:2 ratio. Bulk BCP was performed in 2 iterations to provide a more uniform material removal. Each iteration was conducted using different acid inlet and outlet ports and with the cavity in a different orientation, as depicted in Fig.~\ref{fig:BCP}. The bulk BCP performed on cavity \#1 (\#2) removed an average of 260 (140)~$\mu$m from the bottom subassembly and 180 (160)~$\mu$m from the top subassembly.  

\begin{figure}[!htb]
   \centering
  \includegraphics*[width=250pt]{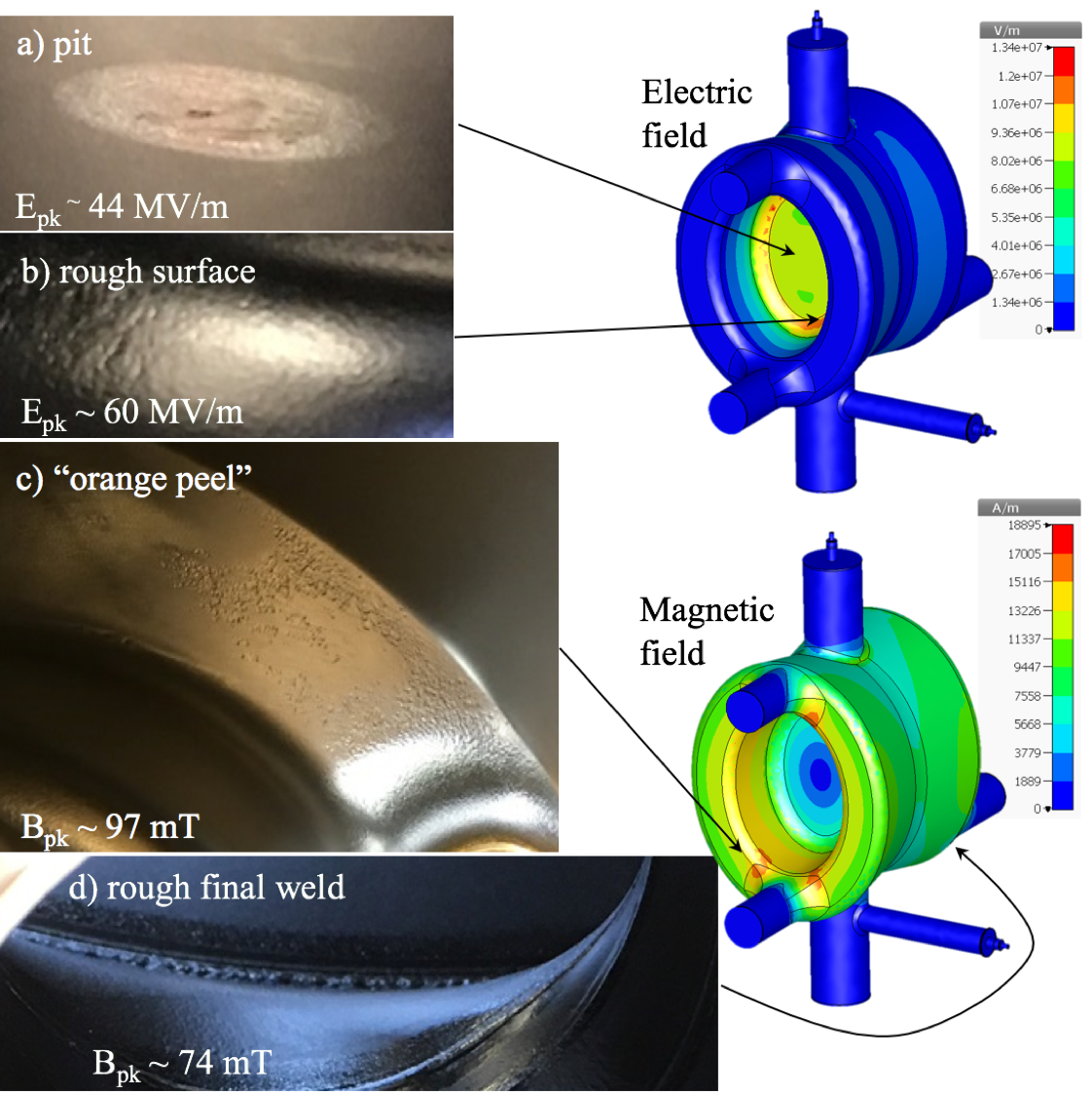}
   \caption{\label{fig:Features}Surface features found in the RF surface of cavity \#1 after bulk BCP. Field values correspond to quench voltage (5.9 MV).}
\end{figure}

The interior surface of cavity \#1 looked smooth after bulk BCP, except for certain features: 1) pits in the center of both capacitive plates, 2) some rough surface in the high electric field region of the capacitive plates, 3) ``orange peel'' or pitting close to the HOM port of the top subassembly and 4) a rough bead in the last weld joining center and bottom subassemblies, in the surroundings of the HOM port which was the closest to the pickup tube. The features are shown in Fig.~\ref{fig:Features}. 

A light BCP of about 40~$\mu$m was performed on the cavities in the same configuration of the first bulk BCP iteration. The cavities were rinsed inside the BCP cabinet multiple times with cold water and then warm water after each BCP iteration. 

\section{Bare cavity cold RF test: cavity \#1}
Cavity \#1 was cold tested in the SRF Facility of Jefferson Lab in February 2017 (the first cryogenic RF test of a DQW SPS-series cavity). The main goals of the test were to: 1) determine the quench limit of the cavity at 2 K and, 2) provide data for future cavity commissioning and operation on multipacting, field emission and heat loads. 

The design and location of test couplers was common to all the DQW SPS-series cavities developed in the US and at CERN. The maximum power available for conditioning was limited to 200~W. The input probe (fixed) was inserted into the short beam port to provide an external Q of about 2$\times 10^{9}$. This value was:  a) low enough for possible multipacting conditioning and, b) high enough to explore the quench limit in case of high-field Q slope.   

\begin{figure}[!htb]
   \centering
  \includegraphics*[width=240pt]{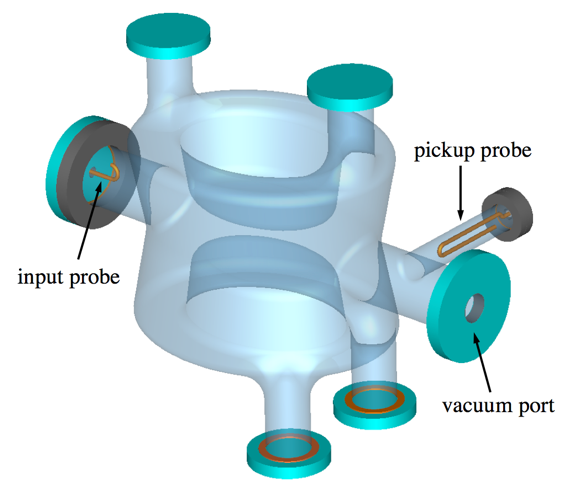}
   \caption{\label{fig:TestPorts}Port distribution for bare DQW cavity cold tests.}
\end{figure}

The pickup probe was inserted into its port for an external Q of about 1$\times 10^{12}$. Both input and pickup probes were hooks made of copper. A sketch depicting the port distribution is shown in Figure~\ref{fig:TestPorts}. The DN100 and DN63 stainless steel flanges had all been coated at CERN with a thin film of niobium, including the zero-length flanges hosting the coupler feedthroughs. Coating the stainless-steel flanges with a niobium film increased the intrinsic Q of these components by six orders of magnitude. RF-seal copper gaskets were used in every DN100 and DN63 flange connections to further reduce power losses~\cite{bib:Shipman16}.  

\begin{figure}[!htb]
   \centering
  \includegraphics*[width=255pt]{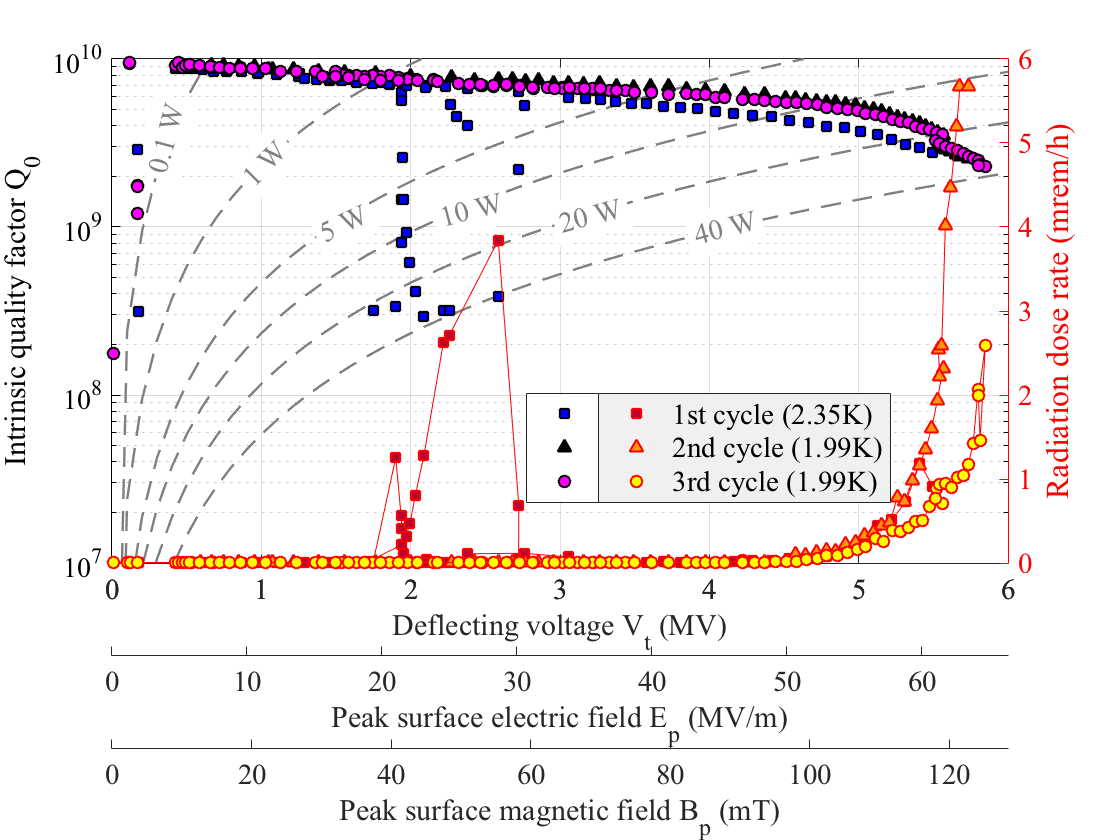}
   \caption{\label{fig:TestResults}$Q_{0}$-$V_{\textrm{t}}$ curves and radiation measured during the cold test of cavity \#1 at Jefferson Lab.}
\end{figure}

The intrinsic Q of the whole cavity assembly was  $(9.8 \pm 1.0) \times 10^{9} $ at low field,  $(6.5\pm 0.6)\times 10^{9} $ at nominal voltage level and $(2.29 \pm 0.17) \times 10^{9}$ at quench field. The cavity reached a maximum deflecting voltage of $(5.9 \pm 0.4)$~MV before quenching, as shown in Fig.~\ref{fig:TestResults}. Errors for these quantities were estimated from the expressions detailed in Ref.~\cite{bib:Powers15}. The calculations assumed a 7\% error coming from cable and reference power meter calibrations and 2\% error from power meter linearity. The VSWR of the RF system setup was 1.12. 

At nominal deflecting voltage (3.4~MV), less than 5~W were dissipated in the cavity. The cavity quenched at 5.9~MV deflecting voltage, when 32~J were stored in the cavity volume with a total RF dynamic loss of 35~W. Iso-power contours are drawn in Fig.~\ref{fig:TestResults}. At quench voltage (5.9~MV), we estimate that around 191~mW were dissipated in the copper input probe, 20~mW in the copper gaskets and 40~mW in the uncoated regions of the beam port flanges. The losses in other components (pickup probe, coated flanges) were less than 0.13~mW.   The cavity walls dissipated about 35~W. The intrinsic Q of assembly components (flanges, gaskets and test probes, without the contribution from the cavity)  was 3.2 $\times 10^{11}$. The surface resistance values used to calculate the dissipated power in each component were 1~m$\Omega$ for copper (the value accounts for anomalous skin effect suffered by good conductors at cryogenic temperatures and 30\% additional losses due to surface roughness), 30~m$\Omega$ for stainless steel and 20~n$\Omega$ for niobium film on the blank flanges. The calculation assumed a constant surface resistance value, independent of the temperature increase due to Joule effect. The surface resistance of the cavity is estimated to be 9~n$\Omega$ from the calculated geometry factor and the measured intrinsic Q at low fields. 

The BCS surface resistance for a niobium cavity operating at frequency $f$  smaller than $10^{12}$ Hz and at a temperature $T$ lower than a half of the critical temperature $T_{c}$ is given by the fitted expression~\cite{bib:PKH11}: 

\begin{equation}
R_{BCS}(\Omega) = 2 \times 10^{-4} \frac{1}{T}\left(\frac{f}{1.5}\right)^2\ exp\left(-\frac{17.67}{T}\right)
\end{equation}

Using this expression, the BCS surface resistance contributes about 1~n$\Omega$ to the estimated surface resistance mentioned above. On the other hand, the contribution of the residual magnetic field to the surface resistance of the cavity is given by~\cite{bib:PKH11}: 
 
 \begin{equation}
 R_{\textrm{mag}} = 0.3(\mathrm{n}\Omega)\ H_{\textrm{ext}}(\mathrm{mOe})\sqrt{f(\mathrm{GHz})}
 \end{equation}
 
where $H_{ext}$ is the residual magnetic field and  $f$ is the operating frequency of the cavity. The residual magnetic field in the Dewar was 5.3~mG maximum, thus contributing with 1~n$\Omega$ to the total surface resistance of the cavity. %{\color{red} Fit with WinSuperfit. During warm up at low field (maximum peak magnetic field below 2.3 mT.  Q factor error of 11\% at low field. Intrinsic Q of cavity (removed contribution from flanges, gaskets and probes). }

Both PoP and SPS cavities exhibited Q-switches at various field levels. Q-switches had been observed in PoP cavity cold tests even after surface reprocessing~\cite{bib:PoP-CERN}. 

\begin{figure}[!htb]
   \centering
  \includegraphics*[width=250pt]{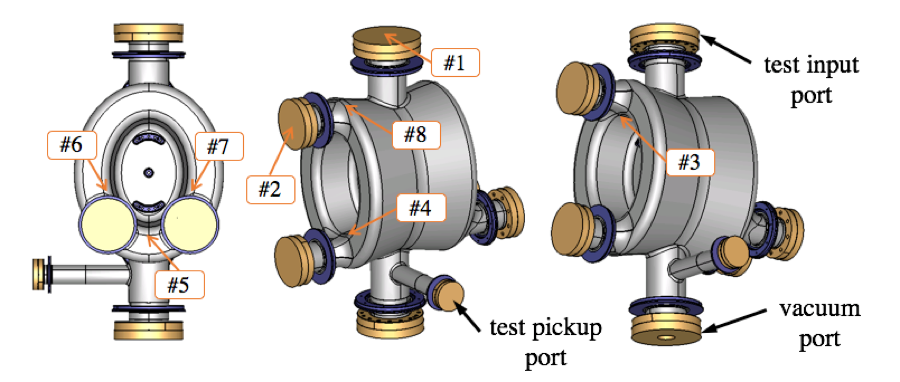}
  \includegraphics*[width=260pt]{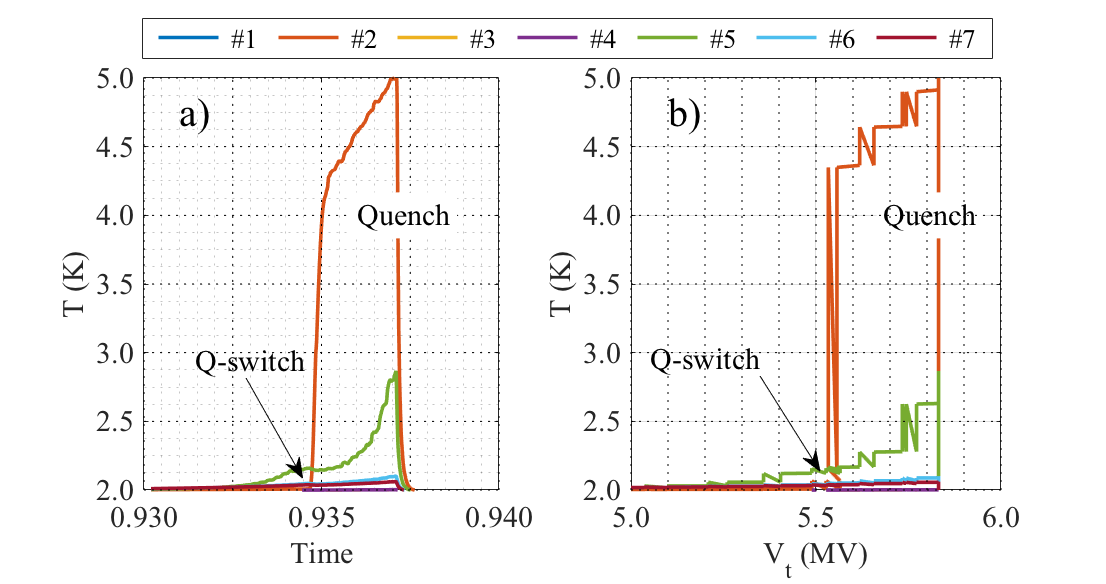}
   \caption{\label{fig:CernoxSignals} [Above] Temperature sensors locations for cold test of cavity \#1 and first cold test of cavity \#2. [Below] Signal recorded from temperature sensors \#1 to \#7 during cold test of cavity \#1: a) temperature versus time; b) temperature versus deflecting voltage.}
\end{figure}

Eight Cernox$^{TM}$ temperature sensors were used to monitor the temperature at critical locations (high field regions, coated flanges, regions with difficult access for cleaning, and the closest point on the cavity to the vapor-liquid helium interface). The location of the temperature sensors is shown in Fig.~\ref{fig:CernoxSignals}. Temperature sensor \#2 $-$ located on top of the FPC port blank flange in the US prototype number 1 $-$ registered an abrupt temperature increase when the cavity had reached a deflecting voltage of $(5.6 \pm 0.2)$~MV. The Q-switch was found at the same voltage level. We suspect that a defect in the niobium coating may be related to the observed Q-switches. 

Temperature sensor \#5 registered a slow temperature increase starting at 5.0~MV deflecting voltage, reaching almost 3~K at quench. Temperature sensor \#5 was located between the two HOM ports in the bottom subassembly. This location corresponds to the region with largest peak magnetic field. The magnetic field in the region is 72.8~mT at 3.4~MV, and 106~mT at 5.0~MV deflecting voltage. The magnetic field is as high as 125~mT at quench voltage in the same region, in contrast to the 0.35~mT reached at temperature sensor \#2.   These field values are those calculated for the ideal cavity model. Temperature sensors \#6 and \#7, located at the other sides of the HOM ports in the bottom subassembly, also registered an increase in temperature but to a lesser extent than temperature sensor \#5. Temperature sensor \#8 did not work appropriately during the test. Signals registered by the temperature sensors are shown in Fig.~\ref{fig:CernoxSignals}. 

The limited time to test this cavity did not allow running in pulsed mode, to investigate if the quench was thermal or magnetic. The FPC port is the shortest port on top of the high magnetic field region of the cavity. The power dissipated in the Nb-coated stainless-steel flange of the FPC port was estimated as 3.38$\times 10^{-6}$~W at quench field. If the niobium film quenched or was damaged, the losses could be as much as 6 orders of magnitude higher and induce a thermal quench. This possibility seems plausible. Temperature sensor \#2 registered a temperature of 5~K right before cavity quench. The temperature difference between surface and bath is larger than the bath temperature itself (at 2~K). In this regime, the heat transfer from the surface into He-II is given by:

\begin{equation}
\dot{q} = \alpha\left( T_{\textrm{s}}^{\textrm{m}} - T_{\textrm{b}}^{\textrm{m}} \right)
\end{equation}

where $\dot{q}$ is the heat flux per unit of area, $T_{\textrm{s}}$ is the surface temperature, $T_{\textrm{b}}$ is the bath temperature. The Kapitza coefficient $\alpha$ and the power-law exponent $m$ are empirical and dependent on material, temperature and surface status. Hereafter the fitted values of $\alpha = 22.4$~W/m$^{2}$/K$^{\textrm{m}}$ and $\textrm{m} = 2.72$ for polished SS304L are used~\cite{bib:Taneda92}. The necessary heat to increase the surface temperature is generated by Joule effect. The power dissipated per unit of area is calculated from:

\begin{equation}
\dot{q} = \frac{1}{2} R_{\textrm{s}}\left|J_{\textrm{s}}\right|^2
\end{equation}
 
where $R_{\textrm{s}}$ is the RF surface resistance of the flange surface exposed to RF and $J_{\textrm{s}}$ is the surface current in the flange at quench field ($J_{\textrm{s}}$ = 480~A/m). The  $R_{\textrm{s}}$ thereby calculated is about a hundredth of an Ohm, closer to the surface resistance of stainless steel than of superconducting niobium and thus supporting the hypothesis of a thermal quench. 
              
On the other hand, the peak surface magnetic field in some regions of the cavity was 125~mT when the cavity quenched, so a magnetic quench cannot be dismissed. In any case, cavities installed in SPS and LHC will not have blank flanges in their ports, so the thermal quench scenario is less likely and higher deflecting voltages could potentially be reached. 

Two main multipacting regions were found during the cold test.  The first region appeared for deflecting voltages lower than 0.3~MV. The PoP cavity also showed this multipacting region and was easily conditioned during the cold tests in 2013 with an adjustable input coupler. Multipacting did not come back once the PoP cavity was conditioned. The limited time available to test the present cavity with a fixed coupler did not allow, however, a full conditioning. 

Simulations predicted weak multipacting signatures in the blending of the small ports for deflecting voltages between 1.7 and 3.0~MV. The impact energy of the electrons multipacting in this region is below 100 eV, for which the SEY coefficient is barely larger than 1. This multipacting band actually appeared between 1.7 and 2.7~MV during the 1st cycle of the cavity test and was easily conditioned in a couple of hours with less than 100 W of RF input power (it does not come back in the 2nd and 3rd cycles). However, the conditioning led to high radiation levels and large Q degradation, as seen in Fig.~\ref{fig:TestResults}. According to simulations, the impact energy of electrons should be small, not enough to make penetrating radiation. In addition, the SEY coefficient is small, so very few electrons are actually stripped from the metal. Therefore, for such a large radiation level and Q degradation, the number of multipacting electrons must also be large and their energy high enough to generate penetrating radiation electron. A ``spray'' of electrons in the wrong phase or initial impact site may travel to cavity regions with higher fields where electrons get accelerated. The impact of these electrons against the cavity walls would then lead to the high radiation observed during the 1st cycle of the test between 1.7 and 2.7~MV. 

Field emission became significant for peak surface electric fields above 45~MV/m, at 4.1 MV deflecting voltage, already larger than nominal. Such high-field onset reflects the good surface quality of the cavity. A reduction of the radiation emitted from the 2nd to the 3rd cycle suggests that the cavity conditioned while measurements were being performed. The maximum peak surface field reached in the cavity during the cold test was about 65~MV/m. The radiation monitor was located inside the shielding hatch, near the Dewar top plate.  

\begin{figure}[!htb]
   \centering 
  \includegraphics*[width=210pt]{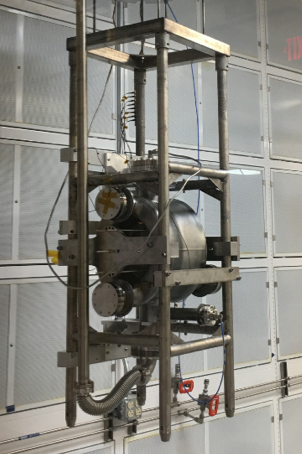}
   \caption{\label{fig:Assembly}Stiffening frame for DQW cold tests at Jefferson Lab.}
\end{figure}

Cavity \#1 and cavity \#2 were tested in vertical orientation. The cavities had a stiffening frame to prevent plastic deformation during cool-down, as shown in Fig.~\ref{fig:Assembly}. The frame was made of titanium. The stiffening frame held the central pin of both capacitive plates. The Lorentz force detuning was -553 Hz/(MV)$^{2}$ for cavity \#1 (6 kHz at nominal deflecting voltage). The pressure sensitivity, measured at low-field during warm up, was -743 Hz/mbar. Lower frequency sensitivities are expected for the dressed SPS-series DQW cavities. The helium vessel and tuning system of the SPS-series DQW cavities~\cite{bib:Zanoni15} should stiffen cavity ports (inductive region) and plates (capacitive region) more than the stiffening frame used for cryogenic RF tests. 

\begin{table*}[htb]
   \centering
   \caption{\label{tab:TestResults}Summary of cavity test performances.}
\begin{ruledtabular}
\begin{tabular}{lcccl}
Magnitude & & Cavity \#1 & Cavity \#2 & Unit \\
 & & (February'17) & (September'17) & \\
 \hline
Maximum deflecting voltage & $V_{\textrm{t}}^{\textrm{max}}$ & 5.9 &  5.3 & MV\\
Maximum stored energy & $U^{max}$ & 32 &  25 & J\\
Maximum dissipated power\footnote{Includes losses in couplers, flanges and gaskets.} & $P_{0}^{\textrm{max}}$ & 35 & 45 & W\\
Maximum peak surface electric field & $E_{\textrm{p}}^{\textrm{max}}$ & 64.7 & 58.2 & MV/m\\
Maximum peak surface magnetic field & $B_{\textrm{p}}^{\textrm{max}}$ &125.2 & 112.6 & mT\\
Intrinsic Q at nominal deflecting voltage & $Q_{0}^{\textrm{nom}}$ & 6.5$\times 10^{9}$ &  7.2$\times 10^{9}$ & \\
Intrinsic Q at maximum field & $Q_{0}^{\textrm{max}}$ & 2.3$\times 10^{9}$ & 1.4$\times 10^{9}$ & \\
Field-emission onset & $V_{\textrm{t}}^{FE}$ & 4.1 & 2.8 & MV\\
% Surface resistance & $R_{\textrm{s}}$ & 9 & &  n$\Omega$ \\
%Pressure sensitivity & $\Delta f/\Delta p$ & -743 & & Hz/mbar \\
%Lorentz force detuning & $\Delta f/\Delta \left(V_{\textrm{t}}\right)^{2}$ &  -449 & & Hz/(MV)$^{2}$ \\
\end{tabular}
\end{ruledtabular}
\end{table*}

\section{Bare cavity cold RF test: cavity \#2}

Cavity \#2 underwent its first cold test in June 2017 at Jefferson Lab. The main goal of this test was to confirm the results obtained with cavity \#1. The cavity presented an intrinsic Q of (9.2 $\pm$ 1.1)$\times 10^{9}$ at low fields and reached up to (5.3 $\pm$ 0.3)~MV deflecting voltage. Several multipacting bands were found and successfully processed in about one hour. The dissipated power was 4~W at nominal deflecting voltage of 3.4~MV. Field emission started at 2.4~MV and led to a pronounced Q-slope. The cavity did not quench; the operation was interrupted due to administrative power limitations. Temperature sensors again identified a temperature increase in the region between the two 45-degree HOM ports when voltages were above 5~MV. 

\begin{figure}[!htb]
   \centering
  \includegraphics*[width=250pt]{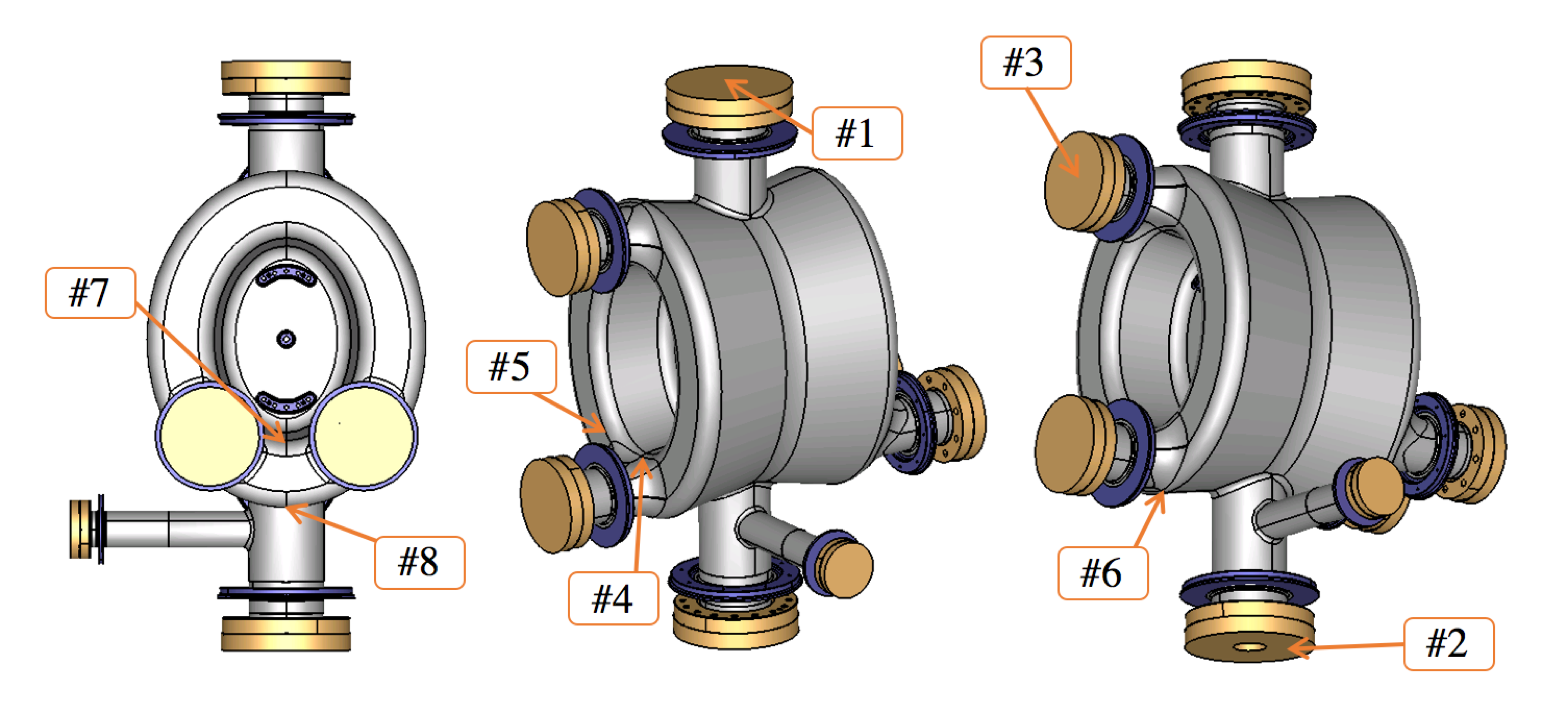}  
  \includegraphics*[width=260pt]{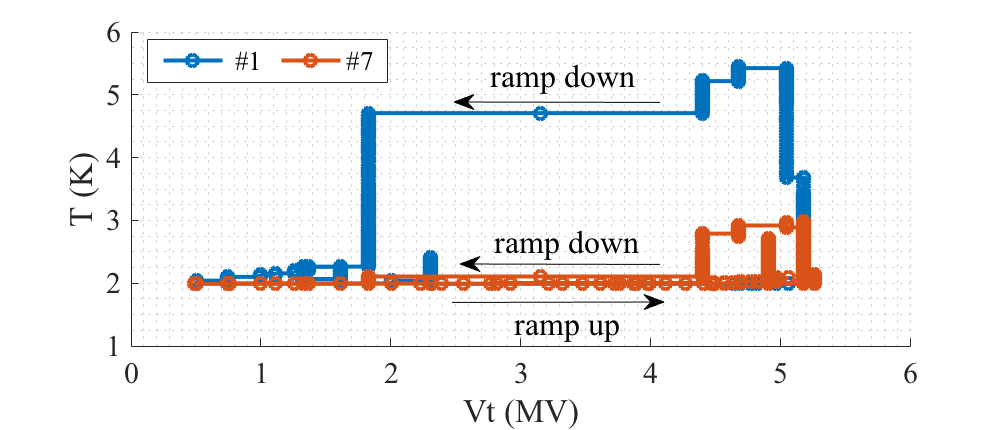}
   \caption{\label{fig:CernoxSignals2} Test of cavity \#2 in September 2017: thermosensor location [above] and signal recorded from thermosensors [below].}
\end{figure}

\begin{figure}[!htb]
   \centering
  \includegraphics*[width=250pt]{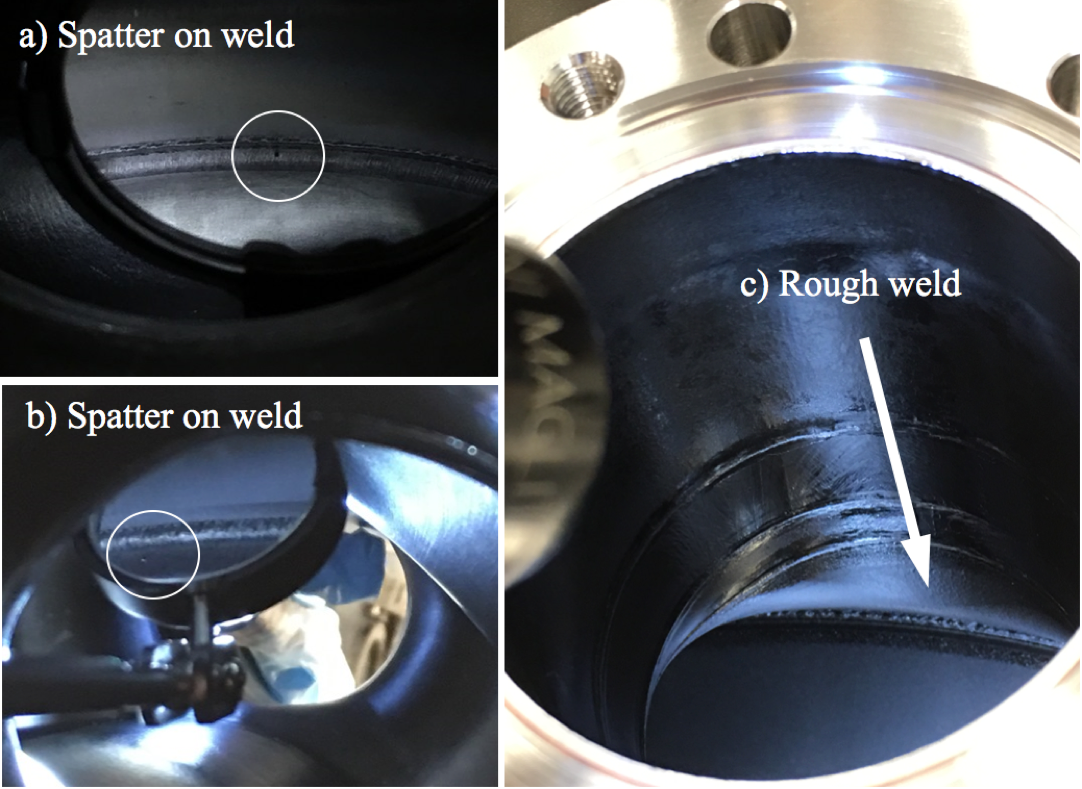}
   \caption{\label{fig:Features2}Surface features in cavity \#2 before bulk BCP. Electric and magnetic fields on the weld area are, respectively, 2 MV/m and 65 mT maximum, at 5.3 MV deflecting voltage.}
\end{figure}

The cavity surface was treated again (light BCP of 18 \textmu m plus HPR) with the aim of increasing the voltage of the field-emission onset and attempting to reach higher deflecting voltages. The second cold test of cavity \#2, also conducted at Jefferson Lab, was held in September 2017. The temperature sensors' distribution for the September test was changed according to Fig.~\ref{fig:CernoxSignals2}. The inspection of cavity \#2 before bulk BCP found some spatter onto and around the two last welds. These welds were also rough. Fig.~\ref{fig:Features2} shows a selection of the features found in the RF surface of cavity \#2. Temperature sensors \#6 and \#8 were placed on top of the location where spatter was observed. Temperature sensors \#4, \#5 and \#7 monitored local and global high magnetic field regions. Temperature sensor  \#3 was placed where previous tests had registered a temperature increase. Temperature sensor \#2 monitored the zero-length flange in the vacuum port where the niobium coating presented a scratch. Lastly, temperature sensor \#1 monitored the saturated helium bath level above the cavity. 

The cavity test found again the low-field (below 0.2 MV), hard multipacting predicted by ACE3P. The cavity was conditioned for 1.5 hours at 10-20 W input power before the first break through. Quenches would later cause the cavity to fall into this low-field multipacting region for about 30 minutes. Soft multipacting regions were found between 1.1 and 3.0 MV and at 4.5 MV, in perfect agreement with CST and ACE3P simulations. About 35 W maximum were used to go through multipacting. 

\begin{figure}[!htb]
   \centering
   \includegraphics*[width=250pt]{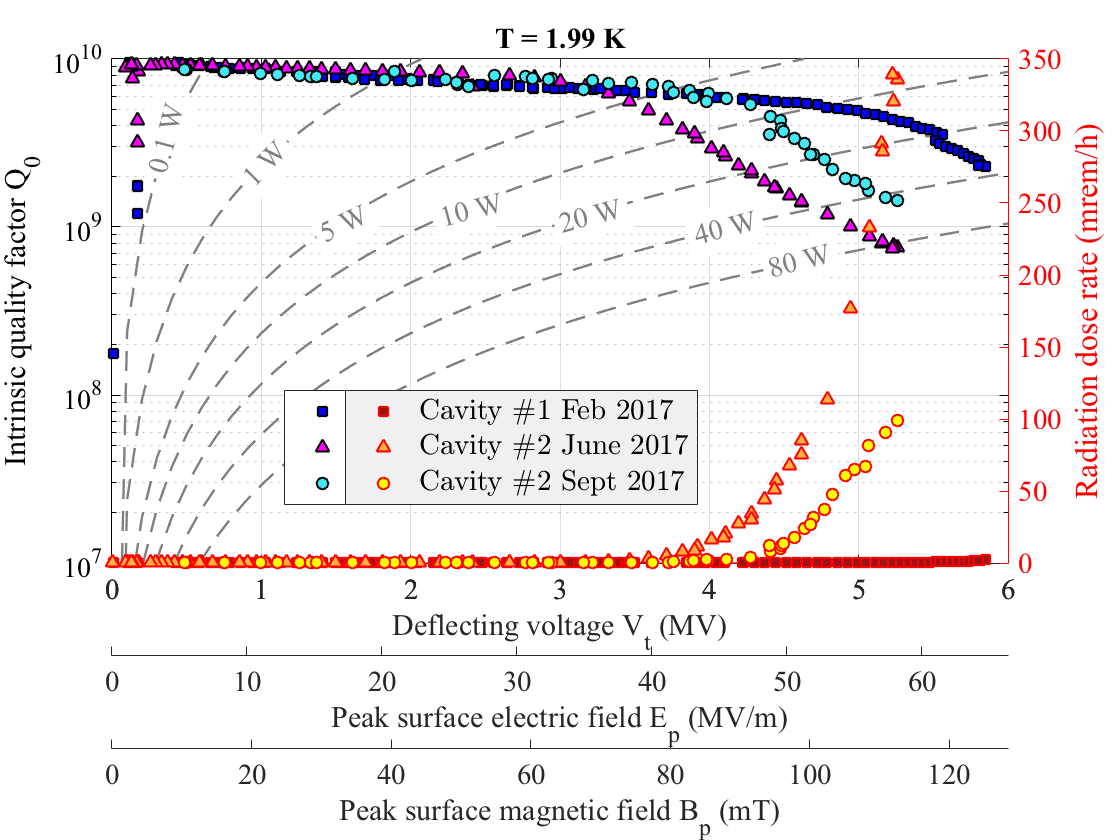}
   \caption{\label{fig:TwoCavityTests}Q-Vt curves measured for the two US-produced SPS-series DQW cavities.}
\end{figure}

The Q-Vt curves for all the bare cavity tests of SPS-series DQW cavities fabricated in the US are displayed together in Fig.~\ref{fig:TwoCavityTests}. In the September test, cavity \#2 quenched at (5.3 $\pm$ 0.2)~MV during CW operation. The same quench limit was found during operation with 1.4~second-long RF pulses at 0.3~Hz repetition rate. This finding suggests that the cavity voltage limitation is due to a magnetic quench. 

The field-emission onset, and consequently the pronounced Q-slope, appeared at higher voltages (2.8~MV) than for the June test. The improvement is attributed to the light BCP and HPR performed between tests. The performance of cavity \#1 and \#2, in terms of voltage, is comparable when accounting for measurement errors. In terms of efficiency, however, the performance of cavity \#2 is not comparable to cavity \#1 due to earlier field-emission (evidenced by a much lower Q and higher radiation at voltages above 4~MV). 

Only temperature sensors \#1 and \#7 registered a signal associated to the quench. Other temperature sensors simply followed the bath temperature, indicating that they were either in a region where temperature did not increase or detached from the cavity surface. After quench, as fields ramped down in the cavity, temperature sensors \#1 and \#7 monitored high temperatures (see Fig.~\ref{fig:CernoxSignals2}). Temperature sensor \#1 reached higher temperature and stayed hot for a longer time after quench than temperature sensor \#7. The longer thermal path from the bath to the niobium coating in the short beam port flange would explain why it took longer time for point \#7 to recover, as the niobium film needs to become superconducting again after a quench. 

\section{Discussion}
Cold tests of the two DQW SPS-series cavities fabricated in US greatly surpassed the nominal deflecting voltage requested per cavity (3.4~MV), going beyond the ultimate deflecting voltage (5~MV) up to 5.3 -- 5.9~MV.  Table~\ref{tab:TestResults} summarizes the cryogenic RF performances of the two bare cavities. The successful results demonstrate the maturity of the DQW cavity RF design. Remarkably, the maximum peak surface fields reached with cavity \#1 are close to those in a typical TESLA-type cavity operating at 30~MV accelerating voltage (60~MV/m and 128~mT)~\cite{bib:TESLA}. The maximum magnetic field of cavity \#1 is comparable to the highest values reached by other SRF cavities that followed a BCP-based surface treatment~\cite{bib:He17, bib:Awida17, bib:Castilla15, bib:Burril10}. The dynamic cryogenic load of the two DQW SPS-series prototypes meets the specification for tests in SPS and operation in LHC. In addition, the test results demonstrate the possibility of manufacturing by industry and the sufficiency of standard SRF surface treatments to reach the required specifications for HL-LHC.  The experience acquired with the design, fabrication and testing of these cavities will serve as a guideline for the development of the eRHIC crab cavities, also based on the DQW concept. 

The test of cavity \#1 exceeded the requested nominal voltage (3.4~MV) by more than 70\% and the ultimate voltage (5.0~MV) by more than 15\%.  The current HL-LHC program envisages the installation of 2 crab cavities per side per beam per IP (from the initial 4 cavities per side per beam per IP). Whereas dressed cavities tend to show worse performances than bare cavities, the test results bring some optimism that 10~MV deflecting voltage can be provided by only 2 cavities while the project envisioned 4 cavities at first. Recent DQW SPS-series cavity tests with a HOM filter have reached 4.7~MV before quench at 2~K and CW operation. Investigations are now in place to push the cavity and filter performance. 

The performance of the two bare DQW SPS-series prototypes was limited by quench. Quenches were accompanied by a large temperature increase in the highest magnetic field region of the cavities and in one of the Nb-coated flanges diametrically opposed to the highest magnetic field region (the short small port in cavity \#1 and the short beam port in cavity \#2). Further tests would be required to determine if the quenches are magnetic or thermal. The maximum achievable voltage may be limited by the use of niobium-coated flanges. Future tests intended to explore the ultimate performance of the DQW cavities should consider the use of niobium extension tubes in replacement of the niobium-coated flanges. 

The limited duration of the tests did not allow for full processing of the multipacting. The DQW cavities will be well overcoupled during the SPS test in 2018 (the fundamental power coupler for the SPS test is designed for an external Q of around 5$\times 10^{5}$), so enough power will be available to process multipacting. Additional tests of bare and dressed cavities will help determining if the origin of Q-switches are coated flanges or the input probe itself. 

The surface features found after the last BCP did not seem to jeopardize the performances of  cavity \#1. On the other hand, cavity \#2 did suffer a pronounced Q-slope accompanied by large radiation. Future studies will evaluate the cavity performances after electropolishing. Electropolishing provides smoother surfaces than BCP~\cite{bib:Saito03}. A reduction of the surface roughness may push the start of the field emission to higher deflecting voltages. 

In the future, we plan to study how nitrogen infusion may impact the cavity performances. Q-slope becomes more acute for fields beyond the nominal deflecting voltage. Nitrogen infusion has twice enabled the state-of-the-art Q at 2~K and a field of  190~mT to be reached in cold tests of 1.3~GHz bulk niobium cavities~\cite{bib:Grassellino17}. The effects of nitrogen infusion in cavities operating at 400~MHz are currently unknown. The nitrogen infusion treatment may have a limited impact on the 400~MHz DQW cavity though. Firstly, recent studies found that low frequency cavities did not show the characteristic $R_{BCS}$ reversal found in 1.3~GHz cavities~\cite{bib:Martinello16}. Secondly, the maximum $Q_{0}$ of the DQW cavities is driven by the intrinsic residual surface resistance.  Studies of flux trapping in the DQW cavity geometry may provide some insight on better cool-down schemes to reduce the magnetic component of the residual surface resistance. The impact on the cavity performance also might be limited for a cavity with such low frequency~\cite{bib:Checchin17}. 

The SPS DQW cavity design satisfies the requirements to be installed in a cryomodule for operation in the LHC. However, the fabrication, preparation and testing of several DQW cavities have shed light on possible design improvements that could boost the cavity performance and/or reduce cavity fabrication time and cost. One first improvement would consist in lengthening the shortest beam pipe to dispense with the niobium coating of the pipe flange. These possibilities are now being studied prior to the production of the LHC-series cavities. 

\begin{acknowledgments}

The authors are thankful to Jingsong Wang (CST) and Chien-Ih Pai (BNL) for technical support to run simulations. Thanks go to Dave Hall and Sam Baurac at Niowave for the careful machining work on the subassembly thinning cuts and RF tuning measurements. We are also grateful to Alick MacPherson (CERN) for providing the niobium coated flanges used in these tests. 

Work partly supported by US DOE through Brookhaven Science Associates LLC under contract No. DE-AC02-98CH10886 and the US LHC Accelerator Research Program (LARP). This research used resources of the National Energy Research Scientific Computing Center, which is as well supported by US DOE under contract No. DE-AC02-05CH11231. The research also received support from the EU FP7 HiLumi LHC - Grant Agreement 284404.  Cavity fabrication at Niowave Inc. was supported by DOE SBIR grant No. DE- SC0007519.

\end{acknowledgments}


\end{document}